\documentclass[useAMS,usenatbib,usegraphicx]{mn2e}
\usepackage{rotating}
\usepackage{txfonts}

\def\lae{\mathrel{<\kern-1.0em\lower0.9ex\hbox{$\sim$}}}
\def\gae{\mathrel{>\kern-1.0em\lower0.9ex\hbox{$\sim$}}}

\title[Number density of ETGs at high-z]
{The number density of superdense early-type 
galaxies at $1<z<2$ and the local cluster galaxies}
\author[P. Saracco et al.]{P. Saracco$^{1}$\thanks{E-mail: 
paolo.saracco@brera.inaf.it}, M. Longhetti$^{1}$,  A. Gargiulo$^{1}$\\ 
$^{1}$INAF - Osservatorio Astronomico di Brera, Via Brera 28, 20121 Milano
}
\begin{document}
\date{Accepted 2010 July 16. Received 2010 June 29; in the original form 2010
April 19}
\pagerange{\pageref{firstpage}--\pageref{lastpage}} \pubyear{2010}
\maketitle

\label{firstpage}

\begin{abstract}
Many of the early-type galaxies observed so far at $z>1$ turned 
out to have smaller radii with respect to that of a typical present-day 
early-type galaxy with comparable mass. 
This has generated the conviction that in the past early-type galaxies were 
more compact, hence denser, and that as a consequence, they should have 
increased their radius across the time to reconcile with the present-day 
ones. 
However, observations have not yet established whether the population 
of early-types in the early universe was fully represented by compact 
galaxies nor if they were so much more numerous than in the present-day 
Universe to require an evolution of their sizes. 
Here we report the results of a study based on a complete sample of 34 
early-type galaxies at $0.9<z_{spec}<1.92$. 
We find a majority (62 per cent) of normal early-type galaxies 
similar to typical local ones, co-existing with compact early-types
from $\sim2$ to $\sim6$ times smaller in spite of the same mass and redshift.
The co-existence of normal and compact early-type galaxies at $<z>\simeq1.5$
{suggests} that their build-up taken place in the first 3-4 Gyr, 
followed distinct paths. 
Also, we find that the number density of compact early-types  at 
$<z>\simeq1.5$ is consistent with the lower limits of the local number density
of compact early-types derived from local clusters of galaxies.
The similar number of compact early-types found in the early and in 
the present day Universe frustrates the hypothesized effective radius
evolution while provides evidence that also compact ETGs were as we see them 
today  9-10 Gyr ago.
Finally, the fact that (at least) most of the compact ETGs at high-z are
accounted for by compact early-types in local cluster of galaxies implies that
the former are the direct progenitors of  the compact early-type 
cluster galaxies establishing a direct link between environment and
early phases of assembly of ETGs.
\end{abstract}

\section{Introduction}
The first studies of scaling relations between effective radius, surface 
brightness and stellar mass performed on the few early-type galaxies 
(ETGs) detected at $z>1$ showed that many of them were smaller than the mean 
local population of ETGs of the same mass (Daddi et al. 2005; di Serego
Alighieri et al. 2005;  Trujillo et al. 2006). 
However, some doubts on this finding were raised since these studies were 
based on Hubble Space Telescope (HST) optical observations sampling the blue 
and UV rest-frame emission dominated by young stars or on 
seeing limited ground-based observations, characteristics which can affect 
the estimate of the effective radius of high-z galaxies. 
The first deep high-resolution (0.075 arcsec/pix) near-infrared ($\lambda=1.6$ 
$\mu$m) 
observations of a sample of 10 ETGs at $1.2<z<1.7$ carried out with the 
Near Infrared Camera and Multi Object Spectrograph (NICMOS)  on-board of HST 
unambiguously showed that the radii of many of them were 2.5-3 times 
smaller than the mean 
radius of local ETGs at comparable mass and surface brightness
(Longhetti et al. 2007). 
Since then, many independent studies found similar results 
(McGrath et al. 2008; Cimatti et al. 2008; van Dokkum et al. 2008;
Damjanov et al. 2009; Muzzin et al. 2009; Cassata et al. 2010; Carrasco et al.
2010)
corroborating the conviction that  high-redshift compact 
ETGs must have increased their radius from their redshift to $z=0$ to 
reconcile their sizes with those of the local ETGs.  
This hypothesis has triggered many theoretical studies defining the 
possible mechanisms able to increase the size of  ETGs: dry (minor) major 
merger in which two spheroids of (non) equal mass merge without involving 
the supposed negligible gas component 
(e.g. Ciotti et al. 2007; Naab et al. 2009; Nipoti et al. 2009); 
adiabatic expansion due to a systematic loss of mass from the central region of the 
galaxy thanks to (e.g.) AGN activity (e.g. Fan et al. 2008); 
age/color (hence M/L) radial gradients which, as the galaxy ages, tend to 
vanish resulting in an apparent increase of the effective radius 
(La Barbera et al. 2009).  
{In principle, any of these models could be tuned to mimic the desired 
effective radius evolution (Hopkins et al. 2010) even if minor dry mergers 
seems to be favoured due to its efficiency in
enlarging the size of galaxies (e.g. Naab et al. 2009) while the other 
effects contributing only for $\sim$20 per cent 
(Hopkins et al. 2010; Bezanson et al. 2009).
However, Nipoti et al. (2009) have recently shown that dry mergers would
introduce substantial scatter in the scaling relations, a scatter not 
observed locally. Consequently, they conclude that ETGs cannot assemble 
more than 40 per cent of their mass via dry mergers.}
Above all, observations have not yet established whether compact ETGs at 
high-z were so much more 
numerous than those in the present-day Universe to require the evolution of 
their effective radius. 
A recent study suggest that not all the high-z early-types were more 
compact than typical local counterparts  and that those more compact were 
older than the others  (Saracco et al. 2009) analogously to what is 
observed in the local 
Universe (Bernardi et al. 2008; Shankar \& Bernardi 2009; Valentinuzzi et al. 2010). 
Thus, the question naturally arising is whether the population of high-z ETGs 
was actually so different from the population we see today,
in particular whether the number density 
of compact ETGs at high-z was so high to require a size evolution
to not exceed the number density of compact ETGs in the local Universe.
To look after this issue we constructed a complete sample of ETGs 
in the redshift range $0.9<z_{spec}<2$ with spectroscopic confirmation of 
their redshift and covered by HST observations.
In this letter we report the analysis we performed and the results
we obtained.
Throughout this paper we use a standard cosmology with
$H_0=70$ Km s$^{-1}$ Mpc$^{-1}$, $\Omega_m=0.3$ and $\Omega_\Lambda=0.7$.
All the magnitudes are in the Vega system, unless otherwise specified.

\section{Sample selection and parameters estimate}
The sample of ETGs we used in this analysis has been selected on the 
southern field of the  Great Observatories  Origins Deep Survey 
(GOODS-South v2; Giavalisco et al. 2004) and it is complete to K$\simeq20.2$.
It has been imaged in four HST-ACS bandpasses (F435W, F606W, F775W and F850LP)
and targeted by extensive observations with ESO telescopes both in the 
optical (3 U-band filters) and in the near-IR (J, H and Ks filters). 
We used the GOODS-MUSIC multiwavelength 
catalog (v.2; Grazian et al. 2006) composed of the 10 photometric bands listed
above and the four Spitzer-IRAC bands 3.6 $\mu$m, 4.5 $\mu$m, 5.8 $\mu$m 
and 8.0 $\mu$m.
The spectroscopic data come from the public ESO-VLT spectroscopic survey
of the GOODS-South field (Vanzella et al. 2008 and references therein).
The sample has been constructed by first selecting all the galaxies brighter 
than K$\simeq$20.2 over the $\sim143$ arcmin$^2$ of the 
GOODS-South field, 
then by removing all the galaxies with measured spectroscopic redshift 
$z_{spec}<0.9$ and those with irregular or disk-like morphology. 
This first step of the morphological classification has been made through
a visual inspection of the galaxies carried out independently by two of us
on the ACS images in the F850LP band.
Finally, on the basis of the best-fitting procedure to the observed 
profile described below, we removed those galaxies (8) having 
a S\'ersic index $n<2$ or clear irregular residuals resulting from the fit.
Out of the 38 early-type galaxies thus selected 34 have measured spectroscopic 
redshift $0.9<z_{spec}<1.92$ leading to $\sim90$ per cent spectroscopic completeness.  

Stellar masses $\mathcal{M}_*$ and ages of the stellar populations were
derived by fitting the last release of the Charlot \& Bruzual 
models (hereafter CB08) to the observed Spectral Energy Distribution (SED) 
of the galaxy at fixed known redshift.
We considered the Chabrier initial mass  function (IMF, Chabrier 2003),  
four exponentially declining star formation histories (SFHs) 
with e-folding time $\tau=[0.1, 0.3, 0.4, 06]$ Gyr and metallicity
0.4 $Z_\odot$, $Z_\odot$ and 2 $Z_\odot$.
Extinction $A_V$ has been fitted in the range $0\le A_V\le0.6$ mag and 
we adopted the extinction curve of Calzetti et al. (2000).
For  $\sim$90 per cent of the sample the best fitting template is defined by SFHs
with $\tau\le0.3$ Gyr, $A_V\le0.4$ mag and $Z=Z_\odot$.
The fact that spectral type and redshift of the galaxies are known
and that their SED is well sampled from the UV to the near-IR rest-frame 
implies that the shape of the best-fitting model, depending on the SFH and
on the age of the model, is sharply constrained. 
In particular, the internal accuracy of our stellar mass estimates is within 
a factor of 1.5 and of age estimate within 0.3 Gyr for age younger than
1.5-2 Gyr and 0.5 Gyr for age older than 2.5 Gyr.
In Fig. 1, the best fitting at fixed redshift to the SED of four ETGs of 
the sample are shown as example.

We derived the effective radius $r_e$ [arcsec] of our galaxies by fitting a 
S\'ersic profile to the observed profile in extremely deep (40-100 ks)
HST ACS-F850LP images using \texttt{Galfit} software 
(v. 2.0.3, Peng et al. 2002).
A PSF to be convolved with the S\'ersic model has been 
constructed for each galaxy by averaging the profile of some bright 
unsaturated stars close to the galaxy itself. 
We point out that the extremely faint limiting surface brightness 
of the ACS observations in the F850LP filter 
($\mu\simeq27.0$ mag/arcsec$^2$) is 5 magnitudes fainter than the faintest 
effective surface brightness 
$\langle\mu\rangle_e^{F850LP}\simeq 22.5$ mag/arcsec$^2$
we measured for our galaxies.
This depth allows an accurate  sampling of the galaxy profiles 
at radii $>2r_e$ assuring the detection of any possible faint wing 
or halo.
{This is a critical point in the estimate of the size of ETGs at high-z
as shown by Mancini et al. (2010).
Indeed, they demonstrate that the lack of high S/N data may result
 in missing a large fraction of the low surface brightness halo
of the galaxy and consequently in underestimating the actual sizes
{ (see also Szomoru et al. 2010)}.
We verified how our estimates compare with those from other studies
based on high S/N data of ETGs at $z>1$ by comparing the average size 
of our ETGs within comparable ranges of stellar masses.
At very high stellar masses there are the studies of 
McGrath et al. (2008) and Mancini et al. (2010) based on the deep HST-NIC2 
images and HST-ACS COSMOS images respectively. 
McGrath et al. studied a sample ($1.4<z<1.6$) in the range 
$1-3\times 10^{11}$ M$_\odot$ characterized by $<R_e>=5.5\pm2$ kpc.
Mancini et al. studied a sample of ETGs ($1.4<z<1.7$)
in the range $2.5-4.5\times 10^{11}$ M$_\odot$ whose average effective radius 
is $<R_e>=7.6\pm5$ kpc.
There are no ETGs more massive than $3\times10^{11}$ in our sample.
However, we estimate $<R_e>=5.7\pm3$ kpc for ETGs in the range 
$2-3\times 10^{11}$ M$_\odot$ and $<R_e>=5.0\pm3$ kpc
in the range $1-3\times 10^{11}$ M$_\odot$.
The ETGs studied by Damjanov et al. (2009) ($1.5\lae z<1.85$)
on deep HST-NIC3 images with masses in the range $0.3-2\times 10^{11}$ 
M$_\odot$ have $<R_e>=2.5\pm1.3$ kpc as those (in the same mass range) 
at $z\simeq1$ studied by Ferreras et al. (2009) on the deep HST-ACS images 
of the GOODS fields (North+South).
Our ETGs in the range $0.3-2\times 10^{11}$ M$_\odot$ have  
$<R_e>=2.6\pm1.7$ kpc.
van der Wel et al. (2005) studied a sample of ETGs on the HST-ACS images of 
the GOODS-South field 12 out of which ($0.9<z<1.15$ and 
masses larger than $0.5\times 10^{11}$ M$_\odot$) are in common with
our sample.
The mean radii of these 12 ETGs resulting from their and our estimate are 
$<R_e>=3.2\pm2.0$ kpc and $<R_e>=3.6\pm2.2$ kpc respectively, in agreement
also with the 12 ETGs at $1<z<1.4$ studied by Newman et al. (2010) in a 
comparable mass range  ($<R_e>=3\pm1.6$).
We have also 8 ETGs in common with the sample studied by Cimatti et al. (2008)
7 out of which in the mass range $2-8\times 10^{10}$ M$_\odot$.
For these ETGs the resulting mean radius is $<R_e>=1.4\pm0.7$ kpc 
from both the studies.
Thus, the average size and scatter we measured for our ETGs agree very well 
with those found in previous studies based on high S/N HST imaging and
spectroscopic redshift.
More difficult is the comparison with other studies focused on the  
population of massive galaxies which usually classify them on the basis
of their SED making use of photometric redshift.
Recently, Williams et al. (2010) studied the size evolution of a large sample
of galaxies selected on the UKIDSS Ultra-Deep Survey (UDS, Warren et al. 2007).
The mean radius resulting from the ground-based K-band images of their 
selected quiescent massive ($>6\times 10^{10}$ M$_\odot$) galaxies is 
$<R_e>\simeq2$ kpc.
This estimate, in agreement with the one derived by Franx et al. (2008) 
on deep ground-based observations of quiescent galaxies in the GOODS-South 
field, is  quite smaller than those reported above for similar mass ranges
($<R_e>\sim3-4$ kpc). 
Similarly, the mean radii ($<R_e><2$ kpc) estimated by Buitrago et al. (2008) 
and Trujillo et al. (2007) on  shallow (2-4 ks) HST images of passive 
galaxies at $z>1.5$ with masses $>10^{11}$ M$_\odot$.
The comparison with these studies is made uncertain both  
by the different methods used to select and classify galaxies and
by the quite different data sets: 
ground-based and/or shallow observations can affect systematically 
the size of galaxies (e.g. Stabenau et al. 2008 and Mancini et al. 
2010) and mass estimate can be affected by the unertainties related to 
SED fitting (e.g. Longhetti \& Saracco 2009; Muzzin et al. 2009; 
Mancini et al. 2010) especially when the redshift is a free parameter 
(Stabenau et al. 2008).
}

\begin{figure}
\begin{center}
\includegraphics[width=8.5cm]{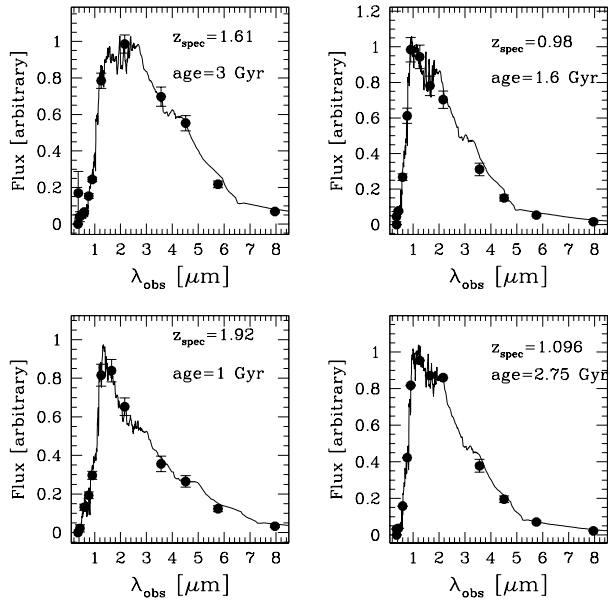} 
\caption{Spectral energy distribution of four early-type galaxies of the
sample. Filled points are the fluxes measured through the 14 photometric
bands U$_{35}$, U$_{38}$, U$_{VIMOS}$, F435W, F606W, F775W, F850LP,
J, H, Ks, 3.6 $\mu$m, 4.5 $\mu$m, 5.8 $\mu$m and 8.0 $\mu$m.
The solid line is the best-fitting template to the observed SED.
}
\end{center}
\end{figure}

\section{Superdense early-types at $1<${\it{\small{z}}}$<2$ and the local 
early-type cluster galaxies}
In Fig. 2 the size-mass (SM) relation, the relation between the 
effective radius R$_e$ [kpc] and the stellar mass $\mathcal{M}_*$ [M$_\odot$] 
of our galaxies (filled symbols), is compared with the relation found by 
Shen et al. (2003) for the local population of ETGs (solid line). 
The original local relation is based on masses derived using 
Bruzual and Charlot (2003) models which provide masses $\sim1.2$ times 
larger than those derived with CB08  models (Longhetti and Saracco 2009).
We thus corrected the relation by this scaling factor to make the
comparison with our data consistent. 
{It is evident that more than half of the sample, 21 out of 34 ETGs 
(filled triangles),
lies within one sigma from the local SM relation, 
that is $\sim$62 per cent of the sample is composed of ETGs, we say normal, 
having morphological and physical parameters which agree with the mean 
local values.}
The recent measure of the velocity dispersion obtained for three normal ETGs,
one of which belonging to the ACS sample (filled triangle marked by open
circle in Fig. 2)  confirms that they are 
similar to the typical local ETGs also from the dynamic point of view
(Cappellari et al. 2009; Onodera et al. 2010). 
{The remaining $\sim38$ per cent (13 galaxies) of the sample is composed of ETGs
(filled circles), we say compact, which diverge more than one sigma from 
the local SM relation having effective radii from $\sim2$ to $\sim6$ 
times smaller than the radius derived from the 
local relation for the same mass.}
This is better shown in the lower panel of Fig. 2 where the degree of 
compactness, defined as the ratio between the effective radius R$_e$
and the effective radius R$_{e,z=0}$ of an equal mass galaxy at $z=0$
as derived from the local SM relation, is plotted as a function of the mass.
The recent attempt to measure the velocity dispersion of a compact galaxy 
(van Dokkum et al. 2009; red open circle in Fig. 2) seems to confirm an 
higher stellar mass density than in normal early-types. 
It is interesting to note that ETGs more compact than the mean local value 
exist for any value of the effective radius and of the stellar mass 
spanned by our sample. 
Moreover,  the degree of compactness does not show any dependence on mass.
Thus, it is clear that the population of high-z ETGs is not primarily 
composed of superdense ETGs but it is dominated by normal ETGs similar 
to local ones. 
{We also report in Fig. 2 the 29 brightest cluster galaxies with 
$\sigma_v>350$ km/s selected by Bernardi et al. (2008) in local cluster 
of galaxies (open squares). 
They populate the SM relation at high masses ($>10^{11}$ M$_\odot$) and
a few of them (5-6)  are compact.} 
\begin{figure}
\begin{center}
\hskip -0.2truecm
\includegraphics[width=8.5cm]{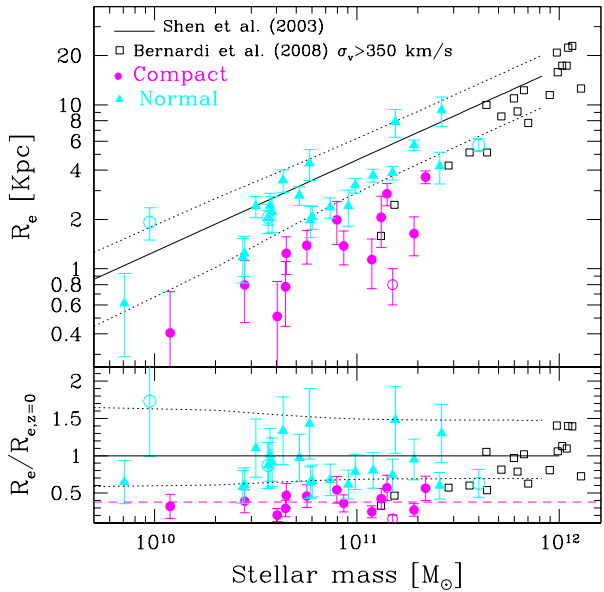} 
\vskip -0.5truecm
\caption{Upper panel - The effective radius
R$_e$ of the 34 ETGs at $0.9<z<2$ is plotted as a function of their 
stellar mass $\mathcal{M}_*$ (filled symbols).
The solid line is the local Size-Mass relation  
$R_e=2.88\times10^{-6}(M_*/M_\odot)^{0.56}$ (Shen et al. 2003) and
the dotted lines represent  the relevant scatter.
Filled circles mark those early-types  more than 1$\sigma$ smaller 
than local ETGs with comparable stellar mass as derived by the local
relation. 
Filled triangles mark the ETGs falling within 1$\sigma$ from the local relation.
Cyan open circles are normal ETGs with  stellar velocity dispersion 
measurements, one of which included in the ACS sample (Cappellari et al. 2009;
Onodera et al. 2010).
Red open circle is the compact ETGs with stellar velocity dispersion 
measurement (van Dokkum et al. 2009).
Open squares are the the brightest cluster galaxies selected by Bernardi et al.
(2008) on the basis of their $\sigma_v>350$ km/s.
Lower panel - The compactness, defined as the ratio between the 
effective radius R$_e$ of the galaxy and the effective radius R$_{e,z=0}$
of an equal mass galaxy at $z=0$ as derived by the local S-M relation, 
is plotted as a function of the stellar mass
Symbols are as in the upper panel.
}
\end{center}
\end{figure}
The critical question is whether the number density of compact ETGs 
in the early Universe was so high to require that they have increased 
their radii to not exceed the number density of compact ETGs in the 
present-day Universe, in practice whether compact ETGs were much more 
numerous at earlier epochs, or if their number has not changed across the time.
A recent estimate by Valentinuzzi et al. (2010) fixes the lower limit of the 
number density of compact ETGs in the local Universe considering compact 
ETGs only in local clusters of galaxies and assuming that no compact 
ETGs are present among the population of non-cluster galaxies.
They find a lower limit of $n=1.8\times10^{-5}$  Mpc$^{-3}$ for stellar 
masses $M_*>3\times10^{10}$ M$_\odot$  and of $n=0.8\times10^{-5}$ Mpc$^{-3}$ 
for masses $M_*>8\times10^{10}$ M$\odot$. 
{Their local densities have been computed considering compact those ETGs 
diverging more than one sigma from the local SM relation of Shen et al. (2003),
 according to our definition 
(Valentinuzzi, private communication).
}
We find that the comoving number density of compact ETGs 
over the volume of about $4.4\times10^5$ Mpc$^3$ sampled by the GOODS area 
between $0.9<z<1.92$ (average redshift $<z>=1.5$) is compatible even with the
local lower limits. 
In particular, we find  $n=(2.2\pm0.7)\times10^{-5}$  Mpc$^{-3}$ for masses 
$M_*>3\times10^{10}$ M$_\odot$ and $n=(0.8\pm0.4)\times10^{-5}$ Mpc$^{-3}$ 
for masses $M_*>8\times10^{10}$  M$_\odot$. 
{van Dokkum et al. (2010) and Guo et al. (2009) indicate that 
Shen et al. (2003) might have underestimated the sizes of massive galaxies by 
a factor of up to two. 
We point out that the method used to define and compare compact ETGs in our 
and in the Valentinuzzi et al. sample (1$\sigma$ from the SM relation) takes 
the results away from any possible systematics in the local relation
since it would affect  the selection of compact ETGs
at high and at low redshift in the same way.}
We computed the co-moving spatial density $n$ and stellar mass density
$\rho$ of early-type galaxies and its statistical uncertainty 
 using the $1/V_{max}$ estimator
where 
\begin{equation}
V_{max}={\omega\over{4\pi}}\int_{z_1}^{z_{max}}{dV\over{dz}}dz
\end{equation}
is the comoving volume.
The solid angle $\omega$ subtended by the 143 arcmin$^2$ 
of the GOODS field is $1.2\times10^{-5}$ strd, $z_1=0.9$ is the minimum 
redshift considered and $z_{max}$ is the maximum redshift at which each 
galaxy would be still included in the sample resulting still brighter 
than K=20.2. 
In Table 1 we summarize our estimates for different ranges of masses and of 
effective radii. 
Given the high spectroscopic completeness (90 per cent) of our sample, the derived 
values could be underestimated by a factor 1.1. 
Moreover, even in the hypothesis that our selection of ETGs was too 
conservative and some of the 8 galaxies removed (see \S 2) because of 
their low S\'ersic index ($n<2$) or their irregular residuals were ETGs, 
the underestimate would not exceed a factor 1.2.
To assess the reliability of our estimates, we compared our number densities 
of ETGs with those in previous studies.
Wuyts et al. (2009) and Bezanson et al. (2009) estimate a number density 
of quiescent galaxies of $n=(12\pm5)\times 10^{-5}$Mpc$^{-3}$  and 
$n=(7\pm2)\times 10^{-5}$Mpc$^{-3}$ for masses $>4\times10^{10}$ M$_\odot$ and 
$>10^{11}$ M$_\odot$ respectively.
Besides the different criterion used to classify galaxies and the use of
photometric redshift, their masses are derived from the Bruzual \& Charlot 
(2003) models with Kroupa IMF which is found to provide 
masses  from 1.2 (Longhetti et al. 2009) to $\sim1.8$ (e.g. Pozzetti et al. 2007; 
Cimatti et al. 2008) times larger than those provided by the CB08 
models with Chabrier IMF. 
Taking into account a scaling factor of 1.5, we estimate a number density of 
ETGs $n=(7\pm2)\times10^{-5}$Mpc$^{-3}$ and $n=(3\pm1)\times 10^{-5}$Mpc$^{-3}$
in the same mass ranges, without considering any correction for
incompleteness. 
Cimatti et al. (2008) estimates a number density $n\simeq10\times10^{-5}$Mpc$^{-3}$
for masses $>10^{10}$ M$_\odot$ while we estimate 
$n=(9\pm2)\times 10^{-5}$Mpc$^{-3}$.
{ We do not try to make a
comparison of the number densities of "compact" ETGs only with other 
studies since it would result extremely uncertain and unreliable.
Indeed, besides the uncertainties discussed above related to the different 
data sets and criteria used to select quiescent or early-type galaxies,
this comparison would be affected also by the different 
criteria used to define "compact" galaxies 
(e.g. R$_e\sim1.5$ kpc and $>8\times 10^{10}$ M$_\odot$, Trujillo et al. 2009;
R$_e\lae1 kpc$ van Dokkum et al. 2009).
}

From the comparison shown in Tab. 1, we see that the number 
density of compact ETGs at high-z is comparable to local values independently
of the mass range and effective radius considered, that is with the number 
density of compact early-types in cluster galaxies. 
Thus, we do not find evidence of an evolution of the number density of 
compact ETGs from $<z>=1.5$  to z=0 and, consequently, of their effective 
radius during the last 9-10 Gyr.

\begin{table*}
\caption{Number and stellar mass density of early-type galaxies in 
the early and in the present-day Universe.
Measurements of the stellar mass density and of the number 
density of early-type galaxies at $<z>=1.5$  are shown for 
the whole population of early-types ($\rho_{<z>=1.5}^{all}$
and $n_{<z>=1.5}^{all}$, left panels) and for the compact early-types 
($\rho_{<z>=1.5}^c$ and $n_{<z>=1.5}^c$, right panels) for different ranges of 
stellar mass $\mathcal{M}_*$ (upper panels) and of effective radius R$_e$ (lower panels).  
$N_{gal}$ is the number of galaxies falling in each interval. 
It is worth noting that a high-density sheet-like structure containing 7 ETGs 
and accounting for an overdensity factor of about 8 in redshift space
is present at $z\sim1.6$ in the GOODS-South field (Cimatti et al. 2008;
Kurk et al. 2009). 
Out of the 7 ETGs five are compact. 
The bracketed values are the estimates obtained without considering the 
5 compact galaxies belonging to this overdensity.
The lower limits to the local number density of compact ETGs $n_{z=0}$ 
are from Valentinuzzi et al. (2010).}
\centerline{
\begin{tabular}{lcccccccc}
\hline
\hline
$\Delta\mathcal{M}_*$ & $N_{gal}^{all}$ & $\rho_{<z>=1.5}^{all}$ & $n^{all}_{<z>=1.5}$ &\vline& $N_{gal}^{c}$ &$\rho_{<z>=1.5}^{c}$ & 
$n^{c}_{<z>=1.5}$ & $n^{c}_{z=0}$ \\
$10^{11}$M$_\odot$ &   & $10^6$ M$_\odot$ Mpc$^{-3}$& $10^{-5}$ Mpc$^{-3}$ &\vline&   &$10^6$ M$_\odot$ Mpc$^{-3}$& $10^{-5}$ Mpc$^{-3}$& $10^{-5}$ Mpc$^{-3}$\\
\hline
$>1$                   & 7 & $1.5\pm0.6$ &$0.95\pm0.4$&\vline&  3(2)& $0.6(0.4)\pm0.4$&0.4(0.2)$\pm$0.2& 0.5 \\
$>0.8$            &12 & $2.3\pm0.7$ & $1.9\pm0.6$&\vline&  5(3)& $1.0(0.6)\pm0.5$&0.8(0.5)$\pm$0.4& 0.8 \\
$>0.3$            &25 & $3.8\pm0.8$ & $5.1\pm1.0$&\vline& 11(7)& $1.6(1.0)\pm0.5$&2.2(1.4)$\pm$0.7& 1.8 \\
$0.1-4$ &33 & $4.6\pm0.9$ & $9.0\pm2.0$&\vline& 13(8)& $1.8(1.1)\pm0.5$&3.4(2.3)$\pm$1.0& 2.4 \\
\hline
\hline
$\Delta R_e$ & $N_{gal}^{all}$ & $\rho_{<z>=1.5}^{all}$ & $n^{all}_{<z>=1.5}$ &\vline& $N_{gal}^{c}$ &$\rho_{<z>=1.5}^{c}$ & 
$n^{c}_{<z>=1.5}$ & $n^{c}_{z=0}$ \\
kpc &   & $10^6$ M$_\odot$ Mpc$^{-3}$& $10^{-5}$ Mpc$^{-3}$&\vline&  &$10^6$ M$_\odot$ Mpc$^{-3}$& $10^{-5}$ Mpc$^{-3}$& $10^{-5}$ Mpc$^{-3}$\\
\hline
$ <1$	&5 &$0.4\pm0.2$ &$0.3\pm0.2$&\vline&4(2)& $0.3(0.2)\pm0.2$ &1.7(1.0)$\pm$1.0&0.6\\
$1-4$&24&$3.4\pm0.7$& $6.4\pm1.0$&\vline&9(6)& $1.5(1.0)\pm0.5$ &1.7(1.2)$\pm$0.6&1.7\\
$2-4$&15&$2.2\pm0.6$& $4.0\pm1.0$&\vline&3(3)& $0.6(0.6)\pm0.4$ &0.5(0.3)$\pm$0.3&0.6\\
\hline
\hline
\end{tabular}
}
\end{table*}

\section{Discussion and conclusions}
The high spectroscopic completeness and the deep HST imaging of the 
GOODS-South field have allowed us to select a complete sample 
of 34 ETGs with K$\leq20.2$ at $0.9<z_{spec}<2$  on the basis of 
their redshift and morphology.
Only 13 of them diverge more than one sigma from the local size-mass
relation having effective radii from $\sim2$ to $\sim6$ times smaller than
the radii derived from the SM relation for the same masses.
Their number density does not exceed the one measured in the
local Universe. 
Indeed, our analysis shows that the young Universe contained nearly the 
same number of compact early-type galaxies than present-day Universe and, 
consequently, that the hypothesis of their growth in radius in the last 9-10 
Gyr is not justified.
On the contrary, this provides evidence that they were as we see them today
even 9-10 Gyr.
{Actually, Mancini et al. (2010) reached similar conclusion for  
very high-mass ($>2.5\times10^{11}$ M$_\odot$) ETGs at $1.4<z<1.7$ and
more recently Newman et al. (2010) show that $<10^{11}$ M$_\odot$ ETGs 
do not show signs of size evolution at $z<1.4$ while estimate a possible 
evolution for larger masses.
It is important to note that these and our results obtained over the redshift 
range $0.9<z_{spec}<2$ imply that if the compact ETGs detected at $z\gae2.3$ 
(e.g. van Dokkum et al. 2008) are actually all so small and represent most
of the population of ETGs at that redshift, then the size evolution 
they should undergo must take place over a very short period of about 1 Gyr.
This would be difficult to account for by mergers in current models 
(see also Newman et al. 2010).} 
The previously claimed size evolution was proposed to explain the apparently 
large number of compact galaxies found in the many of the first samples of
high-z ETGs. 
This enhanced number could be probably due to selection and 
observational biases that remain to be uderstood. 
Recent studies conducted on local samples of galaxies  point 
out that the selection of old galaxies, the criterion usually used 
to select high-z early-type galaxies, is strongly biased toward smaller 
ETGs for fixed mass (Bernardi et al. 2008; Shankar et al. 2009; Valentinuzzi
et al. 2010; Napolitano et al. 2010).
{In addition, the surface brightness dimming acts against the detection of
the extended halos of ETGs resulting in a systematic underestimates of their
sizes when high resolution and high S/N data are not available 
(e.g. Stabenau et al. 2008 and Mancini et al. 2010). 
This could justify why many of the samples of high-z ETGs constructed so 
far are mostly composed of compact ETGs. }
We finally point out that to find a significant (2 sigma at least) excess of 
compact ETGs at high-z with respect to the local lower limits 
we should detect two times more ETGs than what actually detected.  

Also, our analysis
shows that when the Universe was only 3-4 Gyr old 
ETGs fully compatible with typical local ones were the majority 
and, most importantly, co-existed with other ETGs having stellar mass 
densities from 10 to 100 times higher in spite of the same redshift and 
the same stellar mass.
This result, confirming previous finding (Saracco et al. 2009), places 
the origin of the degree of compactness in the first 
3 Gyr of the Universe {suggesting} that compact and normal ETGs are the 
result of different assembly histories. 
We further investigate this issue in a forthcoming paper (Saracco et al. 2010). 
The fundamental questions arising are what is that set out and which is the
assembly history accounting for the composite population of ETGs 
observed at $1<z<2$.
In this regard, it is important to point out the other result we obtained: 
{\it the population of compact ETGs at high-z is nearly accounted for by 
the local population of compact early-type cluster galaxies.
This implies that most of the compact ETGs at high-z are 
the direct progenitors of the local early-type cluster galaxies.}
This  establishes a direct link between environment and compactness
suggesting that the environment can play a major role in the early phase
of assembly of ETGs.
{Finally, it is reasonable to hypothesize that some of them may 
increase further their mass through possible merging events in the last 9-10 
Gyr.
In this case, those already massive at $<z>\simeq1.5$ may transform into
brightest cluster galaxies whose stellar velocity dispersions ($>300$ km/s) 
are compatible with those expected for high-mass compact ETGs at high-z.}

\section*{Acknowledgments}
This work is based on observations made with the ESO telescopes at the Paranal 
Observatory and with the NASA/ESA Hubble Space Telescope, obtained from the 
data archive at the Space Telescope Science Institute which is operated by 
the Association of Universities for Research in Astronomy. 
This work has received financial support from ASI (contract I/016/07/0). 
We are particularly  grateful to Tiziano Valentinuzzi and Bianca Poggianti 
for having provided us with the spatial density of compact early-types in 
local cluster of galaxies computed according to our definition.
We thank Mariangela Bernardi for having provided us with the stellar masses 
of the BCGs with $\sigma_v>350$ km/s.

\section{References}

\noindent  Bernardi, M., Hyde, J. B., Fritz, A., Sheth, R. K., Gebhardt, K., 
Nichol, R. C. A., 2008, MNRAS, 391, 1191

\noindent Bezanson R., van Dokkum P. G., Tomer T., Marchesini D., Kriek D.,
Franx M., Coppi P. 2009, ApJ, 697, 1290

\noindent  Bruzual A.,G. \& Charlot S. 2003, MNRAS 344, 1000

\noindent Buitrago F., Trujillo I., Conselice C. J., Bouwens R. J., Dickinson
M., Yan H. 2008, ApJ, 687, L61

\noindent  Calzetti D., Armus L., Bohlin R. C., Kinney A. L., Koorneef J.,
Storchi-Bergmann T. 2000, ApJ, 533, 682

\noindent  Cappellari M., di Serego Alighieri S., Cimatti A., et al. 2009,ApJ, 704, L34

\noindent  Carrasco E. R., Conselice C. J., Trujillo I. 2010, MNRAS, 405, 2253

\noindent  Cassata P., et al. 2010, ApJ, 714, L79

\noindent  Chabrier G. 2003, PASP, 115, 763

\noindent  Cimatti A., et al. 2008, A\&A, 482, 21

\noindent  Ciotti L., Lanzoni B., Volonteri M. 2007, ApJ, 658, 65

\noindent  Daddi E., Renzini A., Pirzkal N., et al. 2005, ApJ, 626, 680

\noindent  Damjanov I., McCarthy P. J., Abraham R. G., et al. 2009, ApJ, 
695, 101

\noindent  di Serego Alighieri S., et al. 2005, A\&A, 442, 125

\noindent  Fan, L., Lapi, A., De Zotti, G., Danese, L., 2008, ApJ, 689, L101

\noindent Ferreras I., Lisker T., Pasquali A., Khochfar S., Kaviraj S. 2009,
MNRAS, 369, 1573

\noindent  Franx M., van Dokkum P. G., F$\ddot{o}$rster Schreiber N., 
Wuyts S., Labb\'e I., Toft S. 2008, ApJ, 688, 770

\noindent  Giavalisco, M., Dickinson, M., Ferguson, H. C., et al. 2004,
ApJ, 600, L103

\noindent  Grazian A., et al. 2006, A\&A, 449, 951

\noindent Guo Q., White S. 2009, MNRAS, 396, 39

\noindent  Hopkins, P., Bundy, K., Hernquist, L., Wuyts, S., Cox, T. J. 2010,
MNRAS, 401, 1099

\noindent  Kurk J., et al. 2009, A\&A, 504, 331

\noindent La Barbera F., de Carvalho R. R., ApJ, 699, L76

\noindent  Longhetti M., et al., 2007, MNRAS, 374, 614

\noindent  Longhetti M., Saracco P. 2009, MNRAS, 394, 774

\noindent  McGrath E., Stockton A., Canalizo G., Iye M., Maihara T. 2008,
ApJ, 682, 303

\noindent  Mancini, C., Daddi E., Renzini A., et al. 2010, MNRAS, 401, 933

\noindent  Muzzin A., van Dokkum P. G., Franx M., Marchesini D., Kriek M.,
Labb\'e I. 2009, ApJ, 706, L188

\noindent  Naab T., Johansson P. H., Ostriker J. P. 2009, ApJ, 699, L178

\noindent Napolitano N. R., Romanowsky A. J., Tortora C. 2010, MNRAS, 405, 2351

\noindent Newman A. B., Ellis R. S., Treu T., Bundy K. 2010, ApJ, 717, L103  

\noindent  Nipoti C., Treu T., Auger M. W., Bolton A. S. 2009, ApJ, 706, L86

\noindent  Onodera M. et al. 2010, ApJ, 715, L60

\noindent  Peng C.Y., Ho L.C., Impey C.D. \& Rix H-W 2002, AJ 124, 266

\noindent Pozzetti L, et al. 2007, A\&A, 474, 443

\noindent  Saracco P., Longhetti M., Andreon S., 2009, MNRAS, 392, 718

\noindent Saracco P., Longhetti M., Gargiulo A. 2010, MNRAS, submitted

\noindent  Shankar F., Bernardi M. 2009, MNRAS, 396, L76

\noindent  Shen S., et al., 2003, MNRAS, 343, 978

\noindent Stabenau H. F., Connolly A., Jain B. 2008, MNRAS, 387, 1215

\noindent Szomoru  D., et al. 2010, ApJ 714, L244

\noindent  Trujillo I., Feulner G., Goranova Y., et al. 2006, MNRAS, 373, L36

\noindent Trujillo I., Conselice C. J., Bundy K., Cooper M. C., Eisenhardt P.,
Ellis R. S. 2007, MNRAS, 382, 109

\noindent Trujillo I., Cenarro A. J., de Lorenzo-C\'aceres A., Vazdekis A.,
de la Rosa I. G., Cava A. 2009, MNRAS, 692, L118

\noindent Valentinuzzi T., et al. 2010, ApJ, 712, 226

\noindent  van der Wel A., Franx M., van Dokkum P. G., Rix H.-W., Illingworth G. D., 
Rosati P., 2005, ApJ, 631, 145 

\noindent  van Dokkum P. G., et al. 2008, ApJ, 677, L5

\noindent  van Dokkum P. G., Kriek M., Franx M. 2009, Nature, 460, 717

\noindent  van Dokkum P. G., et al. 2010, ApJ, 709, 1018

\noindent  Vanzella E., et al. 2008, A\&A, 478, 83

\noindent  Warren S. J., et al. 2007, MNRAS, 375, 213

\noindent  Williams R. J., Quadri R. F., Franx M., van Dokkum P., Toft S.,
Kriek M., Labb\'e I. 2010, ApJ, 713, 738

\noindent  Wuyts S., et al. 2009, ApJ, 700, 799

\label{lastpage}

\end{document}